\documentclass{article}
\usepackage{graphicx}
\usepackage{booktabs}
\usepackage{tabularx}
\usepackage{makecell}
\usepackage{booktabs}
\usepackage{multirow}
\usepackage{microtype}
\usepackage{ijcai23}
\usepackage{amsmath}
\usepackage{amssymb}
\usepackage{mathtools}
\usepackage{amsthm}
\usepackage{pgfplots}
\usepackage{listings}
\usepackage{amsmath}
\usepackage{pgfplots}
\usepgfplotslibrary{groupplots}
\usepackage{caption}
\usepackage{subcaption}
\usepackage{graphicx}
\usepackage{amsmath}
\usepackage{amssymb}
\usepackage{amsthm}
\usepackage{booktabs}
\usepackage{algorithm}
\usepackage{verbatim}
\usepackage{times}
\usepackage{soul}
\usepackage{url}
\usepackage[hidelinks]{hyperref}
\usepackage[utf8]{inputenc}
\usepackage{graphicx}
\usepackage{amsmath}
\usepackage{amsthm}
\usepackage{booktabs}
\usepackage{algorithm}
\usepackage{algorithmic}
\usepackage[switch]{lineno}

\DeclareMathOperator*{\argmin}{arg\,min}

\title{OSDP: Optimal Sharded Data Parallel for Distributed Deep Learning}

\author{
Youhe Jiang$^1$
\and
Fangcheng Fu$^1$\and
Xupeng Miao$^{2}$\and
Xiaonan Nie$^1$\and
Bin Cui$^{1,3}$
\affiliations
$^1$School of CS \& Key Lab of High Confidence Software Technologies (MOE), Peking University\\
$^2$Computer Science Department, Carnegie Mellon University\\
$^3$Institute of Computational Social Science, Peking University (Qingdao)\\
\emails
youhejiang@gmail.com,
\{ccchengff, xupeng.miao, xiaonan.nie, bin.cui\}@pku.edu.cn,
}
\begin{document}
\usetikzlibrary{plotmarks}
\usetikzlibrary{patterns}
\maketitle

\begin{abstract}
Large-scale deep learning models contribute to significant performance improvements on varieties of downstream tasks.
Current data and model parallelism approaches utilize model replication and partition techniques to support the distributed training of ultra-large models.
However, directly deploying these systems often leads to sub-optimal training efficiency due to the complex model architectures and the strict device memory constraints.
In this paper, we propose Optimal Sharded Data Parallel (OSDP), an automated parallel training system that combines the advantages from both data and model parallelism.
Given the model description and the device information, OSDP makes trade-offs between the memory consumption and the hardware utilization, thus automatically generates the distributed computation graph and maximizes the overall system throughput.
In addition, OSDP introduces operator splitting to further alleviate peak memory footprints during training with negligible overheads, which enables the trainability of larger models as well as the higher throughput. 
Extensive experimental results of OSDP on multiple different kinds of large-scale models demonstrate that the proposed strategy outperforms the state-of-the-art in multiple regards. Our code is available\footnote{\url{https://github.com/Youhe-Jiang/IJCAI2023-OptimalShardedDataParallel}}.
\end{abstract}

\section{Introduction}

Large-scale deep learning (DL) models have achieved great success in the last few years. For example, the pre-trained models, such as ELMo, GPT-3, and LLaMA~\cite{devlin2018bert,raffel2019exploring,kaplan2020scaling,swintransformer,llama}, achieve significant accuracy gains with the explosion of the number of model parameters~\cite{shoeybi2020megatronlm,chatgpt,radford2019language,brown2020language,peters2018deep,lin2021m6}. However, how to load large models into the limited device (e.g., GPU) memory and perform efficient training remains a huge challenge. For instance, none of existing single GPU devices could accommodate a Transformer-based GPT-3 model with 175 billion parameters without involving model distillation or compression techniques. Therefore, building an efficient distributed training system is becoming increasingly important and indispensable for the advanced exploration of deep learning approaches.

% can hardly fit into one single GPU, not to mention train it efficiently. Therefore, distributed model training plays an increasingly important and indispensable role in nowadays large-scale model training.

There are a variety of literature on distributed training methodologies such as data parallel \cite{dean2012large,shallue2018measuring,zinkevich2010parallelized}, model parallel, pipeline parallel \cite{harlap2018pipedream,huang2019gpipe,narayanan2021memory,miao2023sdpipe,yang2020pipemare,nie2023flexmoe} and so on.
Data parallelism accelerates the model training by making each device to be responsible for only a fraction of the input data. However, it requires each device to hold a whole model replica during the training process and collaborate with each other through model synchronizations. Apparently, such a redundant model storage does not resolve the memory bottleneck per device. Model parallelism and pipeline parallelism are promising research directions. For example, Megatron-LM~\cite{shoeybi2020megatronlm} partitions the model parameters and computation in each layer to multiple devices. But they also bring significant inter-device communications on the intermediate results and lead to unacceptable training efficiency. 
Recently, Zero Redundancy Optimizer (ZeRO) has been proposed to eliminates the memory redundancies while retaining low communication overheads.
It only partitions the model parameters to reduce the memory usage but remains the data parallel computation through sharding and gathering model states across the devices. There are several popular implementations, such as DeepSpeed~\cite{rajbhandari2020zero}, AngelPTM~\cite{nie2023angel} and Fully Sharded Data Parallel (FSDP) in FairScale~\cite{baines2021fairscale}, and the latter has been integrated into PyTorch~\cite{paszke2017automatic}. These systems have been successfully used in producing large pretrained models in real industrial scenarios like Microsoft and Meta.

% While data parallelism realizes efficient training by invoking multiple devices to read and process data simultaneously, it does not minimize the memory footprint of model training, which results in its no support for many large-scale model training tasks. Model and pipeline parallelism split a complete model into multiple pieces and process them separately, while provide support for larger-scale model training, communications between operators or layers become system bottlenecks and lead to unacceptable training efficiency. In this case, recent approaches incorporate memory optimization technique into data parallelism. This technique decreases peak memory usage and supports larger models in data parallel training by sharding and gathering model states across data parallel workers, which plays a vital role in today's distributed training systems as well as hybrid parallel strategies. There are several popular implementations such as Zero Redundancy Optimizer (ZeRO) developed by Deepspeed \cite{rajbhandari2020zero} and Fully Sharded Data Parallel (FSDP)\cite{baines2021fairscale} developed by Fairscale.

However, these ZeRO-based systems have two major defects: (1) The zero memory redundancy (i.e., all model parameters are sharded) target is overambitious and brings an additional 50\% communication overhead compared to vanilla data parallel, resulting in a great hardware efficiency reduction. (2) The gigantic tensors involved during the training process may cause peak memory usage beyond the device's memory capacity.
To address the above problems, in this approach, we propose an novel distributed training system Optimal Sharded Data Parallel (OSDP) to achieve a better trade-off between the memory consumption reduction and the training efficiency improvement. Specifically, OSDP breaks the memory redundancy minimization limitation in previous ZeRO-based systems and enables to determine whether to perform parameter sharding for each operator individually. Moreover, to avoid the gigantic tensors (e.g., \texttt{MatMul} outputs), OSDP supports operator splitting and fine-grained memory management, enlarging the entire decision space. Given the specific model description and device information, OSDP provides an efficient search engine to automatically find the optimal parallel strategies for each operator in the computation graph and eventually generates the execution plans. In general, OSDP provides both flexibility and universal applicability to large model distributed training as well as maximizes training efficiency.

We demonstrate the flexibility and efficiency of OSDP on GPT-like Transformers with varying scales and architectures under device memory constraints of 8G and 16G, respectively. Experimental results show that OSDP improves the overall training throughput by up to $2.84\times$ compared with the state-of-the-art parallel training systems.
% outperforms the state-of-the-art parallel training systems in multiple regards, and achieves up to $2.84\times$ the overall system throughput speedup.

\section{Background and Related Works}
\label{sec:bg}

\begin{figure}[!t]
\includegraphics[width=1.0\linewidth]{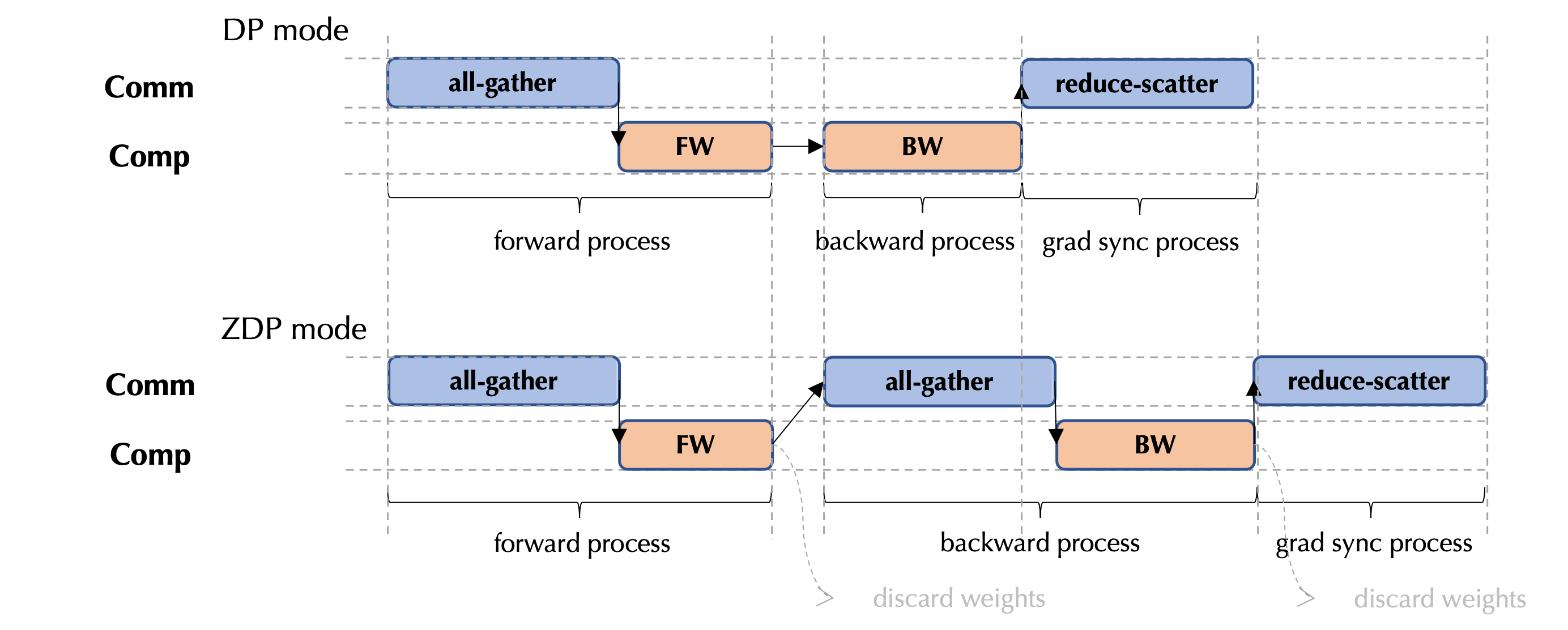}
\caption{The gantt chart of processing one operator in the Data Parallel (DP) mode and the ZeRO Data Parallel (ZDP) mode, respectively.}
\label{fig:DPvsZDP}
\end{figure}

\subsection{Data Parallel}\label{sec:DP}
With the ever-increasing data volume in modern deep learning tasks, data parallelism has become one of the most popular distributed training schemes. 
% It improves training efficiency by enabling multiple processes in multiple devices to read and process data simultaneously, which in case solves the problem that the task overhead is unacceptable due to the slow processing of data by a single device. 
% Data parallelism in deep learning tasks contains three stages. In the model initialization stage, the model parameters on a single device need to be broadcast to all other devices to ensure that each device possesses complete model parameters, which provides them with the ability to maintain a complete forward and backward propagation. During the training stage, 
When training in data parallel, each process maintains a full copy of the model parameters. In each iteration, each device reads in a mini-batch from different data shards to perform the forward and backward propagation accordingly. Then, model gradients are synchronized among the data parallel processes to update the model parameters. In the era of deep learning, the most commonly used technique for gradient synchronization is the all-reduce collective communication operation. In this work, to ease the analysis of communication cost, we follow the previous studies~\cite{rajbhandari2020zero,baines2021fairscale} to dissect an all-reduce operation into a reduce-scatter operation and an all-gather operation, as shown in Figure~\ref{fig:DPvsZDP}.
% the devices that complete the computation earlier in an iteration wait for the slower. In the gradient synchronization stage, all devices complete the communication and summation of gradients through the collective communication operator all-reduce.

\subsection{Zero Redundancy Data Parallel}\label{sec:ZDP}
As the model becomes larger, data parallelism can hardly afford the memory required to maintain the full copies of model parameters, gradients, and optimizer states on each device. Thus, Zero Redundancy Optimizer (ZeRO) is incorporated into data parallelism, well-known implementations include DeepSpeed and Fully Sharded Data Parallel (FSDP) proposed by FairScale and the latter has been integrated into PyTorch \cite{rajbhandari2020zero,baines2021fairscale}.

For ZeRO-based systems such as FSDP, the model parameters, gradients, and optimizer states are partitioned according to the data parallel processes. Each process only stores and updates part of the models. The workflow of one iteration is demonstrated in Figure~\ref{fig:DPvsZDP}. During the training stage, all-gather operations are performed among all processes to get the fully updated parameters, reduced gradients and parameters outside its partition to confirm the complete forward and backward propagation of the model, which greatly reduces the memory consumption of model training while incurs $1.5\times$ communication overhead. During gradient synchronization stage, reduce-scatter operations are performed to synchronize gradients on each device. ZeRO-based systems significantly eliminate model state redundancies during data parallel training, the memory consumption of model states is reduced to $\frac{1}{N}$ compared with before, which ensure continuously efficient training of large-scale models under limited GPU resources.

\subsection{Checkpointing}

Checkpointing \cite{chen2016training,jain2020checkmate,nie2022tsplit} is a widely-used technique to reduce the memory footprint of intermediate activations during training with additional recomputation, which trades roughly 30\% additional computation cost with linear memory allocation. Many recent efforts have been made to incorporate checkpointing with ZeRO data parallelism for better memory management \cite{rajbhandari2020zero,ren2021zerooffload,rajbhandari2021zeroinfinity}. However, when checkpointing is used in ZeRO-based systems (such as DeepSpeed, FairScale, and PyTorch), an additional round of communication is required for the recomputation phase since model parameters are sharded across data parallel processes. 
% DeepSpeed, PyTorch, FairScale and et al. all provide efficient realization of checkpointing.

% checkpointing is usually integrated with the ZDP mode in many real-world applications \cite{rajbhandari2020zero,ren2021zerooffload,rajbhandari2021zeroinfinity}.

\subsection{Other Parallel Training Strategies}
In addition to data parallelism, model parallelism also plays an important role in large-scale model training. There are numerous model parallelism approaches targeting at reducing the memory cost while sustaining training efficiency. For example, Tensor Parallelism (TP)~\cite{shoeybi2020megatronlm} divides the model tensors into multiple parts and train them separately on different devices, which greatly reduces the memory cost while incurs frequent communication during training. Pipeline Parallelism (PP)~\cite{huang2019gpipe,harlap2018pipedream}, different from TP, treats the model as a sequence of layers and partitions them into multiple stages across devices to minimize the memory cost, while consistent communication of the intermediate results is necessary to complete the network propagation. Moreover, recent approaches demonstrate that better performance could be achieved in distributed training with the combination of different parallel strategies. For example, PipeDream \cite{harlap2018pipedream} adopts data parallelism to duplicate the pipeline stages and maximize the system throughput during pipeline parallel training. And a lot of frameworks such as DeepSpeed \cite{rasley2020deepspeed} and Hetu~\cite{DBLP:journals/chinaf/MiaoXP22,Miao_2022,miao2021het} provide efficient realization of 3D parallelism (a combination of data, tensor and pipeline parallelism) for large model training.

\section{OSDP: Optimal Sharded Data Parallel}

In this section, we first formulate the searching problem of execution plans, and then introduce the proposed Optimal Sharded Data Parallel (OSDP) framework.
We first present the frequently used notations:

\begin{itemize}
    \item $N$: the parallelism degree;
    \item $n$: the number of operators in the DL model;
    \item $b$: the training batch size;
    \item $p_i$: the parallel mode of the $i$-th operator;
\end{itemize}

\subsection{Motivation and Problem Formulation}
\label{sec:formulation}

\textbf{\textit{Motivation.}}
As introduced in Section~\ref{sec:bg}, Data Parallel (DP) mode and Zero redundancy Data Parallel (ZDP) mode have distinct memory consumption and communication load --- as shown in Figure~\ref{fig:DPvsZDP}, ZDP consumes fewer memory by discarding model states at the cost of extra communication for re-gathering them. Motivated as such, we wish to utilize the trade-offs between the memory consumption reduction and the training efficiency improvement, and eventually maximize the overall system throughput.

% In general, our goal is to make trade-offs between the memory consumption reduction and the hardware efficiency improvement, which maximizes the system throughput. It provides two modes for processing each operator, including Data Parallel (DP) and Zero redundancy Data Parallel (ZDP). As shown in Figure~\ref{fig:DPvsZDP}, they have distinct communication patterns.
% During the model initialization and forward processes, both the DP and ZDP operators shard and all-gather their parameters across data parallel process. Full weights of ZDP operators are discarded for memory saving after forward pass. During the backward process, DP operators compute their gradients directly, while ZDP operators all-gather full weights across data parallel process, compute their gradients and then discarded full weights again at the end of backward pass. During gradient synchronization process, both DP and ZDP operators perform reduce-scatter operations across data parallel process to synchronize gradients.

To achieve this goal, we develop Optimal Sharded Data Parallel (OSDP), a parallel training framework that guarantees training efficiency as well as breaks the memory bottlenecks of existing data parallel implementations. Given a DL model, OSDP automatically searches for the optimal execution plan that maximizes training throughput while satisfying the device memory limits. %In particular, for each operator in the model, OSDP measures the memory and time cost under DP and ZDP modes. 

\textbf{\textit{Problem Formulation.}}
To be formal, we formulate the optimal execution plan searching problem as follows.
Given a DL model with $n$ operators, where each operator is processed in either DP or ZDP mode, assuming the available device memory is denoted as ${M}\_{limit}$, OSDP searches for the optimal parallel modes for all operators $\boldsymbol{p} = \{p_i\}_{i=1}^{n}$ and the corresponding training batch size $b$ to minimize the averaged training time (i.e., maximize the overall training throughput):
\begin{equation}
\begin{aligned}
\boldsymbol{p}^*, b^* &= \argmin_{\boldsymbol{p}, b} T(\boldsymbol{p}, b)
\coloneqq \frac{1}{b} \sum_{i=1}^{n} T_i(p_i, b)
\\
\text{ s.t. } 
&M(\boldsymbol{p}, b) \coloneqq \sum_{i=1}^{n} M_i(p_i, b) \leq {M}\_{limit}, \\
& p_i \in \{DP, ZDP\} \text{ for } i \in \{1, 2, ..., n\}, \\
& b \in Z^+,
\end{aligned}
\end{equation}
where $M_i(p_i, b), T_i(p_i, b)$ denote the memory and time cost of the $i$-th operator when training in the parallel mode of $p_i$ and with a batch size of $b$. 
Then, we introduce how OSDP estimates the memory and time cost, respectively.

To estimate the \textit{memory cost}, we take three types of data into account: model states (including model parameters and optimizer states), intermediate activations, and the extra overhead (such as the temporary workspaces required by the operator). For simplicity, for the $i$-th operator, the memory consumed by these three types of data are denoted as three factors $M_i^{(model)}, M_i^{(act)}, M_i^{(extra)}$, respectively. Since ZDP shards the model states across the parallel processes, the memory consumption of model states could be amortized to $1/N$, where $N$ is the number of parallel processes. Therefore, the memory cost can be expressed as 
\begin{equation*}
\begin{aligned}
\small
M_i(p_i, b) = 
    \begin{cases}
        M_i^{(model)} + b M_i^{(act)} + M_i^{(extra)}, &\text{if } p_i \text{ is } DP \\
        \frac{M_i^{(model)}}{N} + b M_i^{(act)} + M_i^{(extra)}, &\text{if } p_i \text{ is } ZDP
    \end{cases}.
\end{aligned}
\end{equation*}
It is worthy to note that although the memory factors (i.e., $M_i^{(model)}, M_i^{(act)}, M_i^{(extra)}$) vary for different operators, they can be calculated according to the definition of operators (e.g., types and shapes). Consequently, after the \textit{model description} is provided, OSDP immediately computes the memory factors for the searching of execution plans.

% Denote \red{xxx}, the memory cost can be modeled as follows ($m_z$ for ZDP operators and $m_d$ for DP operators):
% \begin{equation}
%     \begin{split}
%     m_z = \frac{model\_states}{N} + activations * bsz + extra\_overhead \\
%     m_d = model\_states + activations * bsz + extra\_overhead
%     \end{split}
% \end{equation}

The \textit{time cost} consists of communication and computation cost, where the communication cost is related to the amount of model parameters, while the computation cost is related to the training batch size. Thus, we model the communication and computation cost through the $(\alpha,\beta,\gamma)$-model \cite{hockney1994communication,thakur2005optimization,2021}, where $\alpha$, $\beta$, $\gamma$ represent the network latency, transfer time per byte, and computation coefficient, respectively. 

To model the communication cost, we follow the ring-based all-gather and reduce-scatter operations as supported by NVIDIA Collective Communication Library (NCCL) \cite{chan2007collective}. To accomplish one all-gather or reduce-scatter operation, $N-1$ communication steps are required and the amount of communication in each step is ${S_i}/{N}$, where $N$ is the number of parallel processes and $S_i$ is the size of model parameters for the $i$-th operator. As depicted in Figure~\ref{fig:DPvsZDP}, if an operator is processed in ZDP mode, three collective operations (two all-gather operations and one reduce-scatter operation) are needed, resulting in $3(N-1)$ communication steps, while $2(N-1)$ communication steps are expected for DP mode. As for the computation cost, it is proportional to the training batch size and the computation coefficient. Putting them together, the time cost can be modeled as 
\begin{equation*}
\begin{aligned}
\small
T(i; p_i, b) = 
    \begin{cases}
        2(N-1)(\alpha + \frac{S_i}{N}\beta) + b\gamma_i, &\text{if } p_i \text{ is } DP \\
        3(N-1)(\alpha + \frac{S_i}{N}\beta) + b\gamma_i, &\text{if } p_i \text{ is } ZDP
    \end{cases}.
\end{aligned}
\end{equation*}
Similar to the memory factors, the size of model parameters (i.e., $S_i$) can be calculated via the model description. However, the values of $\alpha$, $\beta$, $\gamma_i$ vary according to hardware ability, experimental environments, and operator types. In practice, we require that such \textit{device information} has been profiled in advance and is provided for the optimal plan searching in OSDP. 
Our problem formulation does not consider the overlapping between communication and computation, as the communication cost usually dominates in large model training.

% \jyh{However, since the communication cost usually dominates in large model training, the communication and computation overlap is not considered in our problem formulation.}

% , so each of them should be profiled before the modeling of the cost. As a result, the Profiler outputs the memory and time cost of each operator with different processing modes, as well as the device memory limit obtained from the input device information.

\begin{algorithm}[!t]
\caption{Routines of OSDP.}
\label{alg:CPSA}
\textbf{Input}: Model Description $MD$, Device Information $DI$. \\
\textbf{Output}: The optimal execution plan $\boldsymbol{p}^*$ and the corresponding batch size $b$. 

\begin{algorithmic}[1] %[1] enables line numbers
\STATE Initialize candidate plans $\mathcal{P} \leftarrow \{\}$
\STATE \textit{// Iteratively increase the training batch size.} 
\FOR{training batch size $b \in \left\{1, 2, 3, ...\right\}$}
\STATE $T^*(b) \leftarrow \textup{INF}, \boldsymbol{p}^*(b) \leftarrow \textup{None}$
\STATE \textit{// Traverse execution plans via Depth First Search.} 
\FOR{execution plan $\boldsymbol{p} \in \left\{DP, ZDP\right\}^n$}
\STATE Estimate memory and time cost $M(\boldsymbol{p}, b), T(\boldsymbol{p}, b)$
\IF{$M(\boldsymbol{p}, b) \leq {M}\_{limit}$ and $T(\boldsymbol{p}, b) < T^*(b)$}
\STATE $T^*(b) \leftarrow T(\boldsymbol{p}, b), \boldsymbol{p}^*(b) \leftarrow \boldsymbol{p}$
\ENDIF
\ENDFOR
\IF{$\boldsymbol{p}^*(b)$ is $\textup{None}$}
\STATE \textit{// Stop searching since all plans exceed memory limit \\// under the current batch size.}
\STATE break
\ELSE
\STATE $\mathcal{P} \leftarrow \mathcal{P} \cup \left\{(T^*(b), \boldsymbol{p}^*(b), b)\right\}$
\ENDIF
% \STATE \textit{// Generate plan and throughput.}
% \STATE plan, throughput = Search\_Engine(time\_cost, memory\_cost);
% \STATE Cand\_throughput.append(throughput), Cand\_plan.append(plan);
% \STATE // Stop while all plans exceed memory limit.
% \IF{plan == None}
% \STATE break
% \ENDIF
\ENDFOR
\STATE \textit{// Return the optimal execution plan $\boldsymbol{p}^*$ and the \\// corresponding batch size $b^*$.}
\STATE $\boldsymbol{p}^*, b^* \leftarrow \argmin_{\boldsymbol{p}, b} \left\{ T(b) \vert (T(b), \boldsymbol{p}, b) \in \mathcal{P} \right\}$
\STATE \textbf{return} $\boldsymbol{p}^*, b^*$
\end{algorithmic}
\end{algorithm}

\subsection{System Overview}

\begin{figure}[!t]
\centering
\includegraphics[width=1\linewidth]{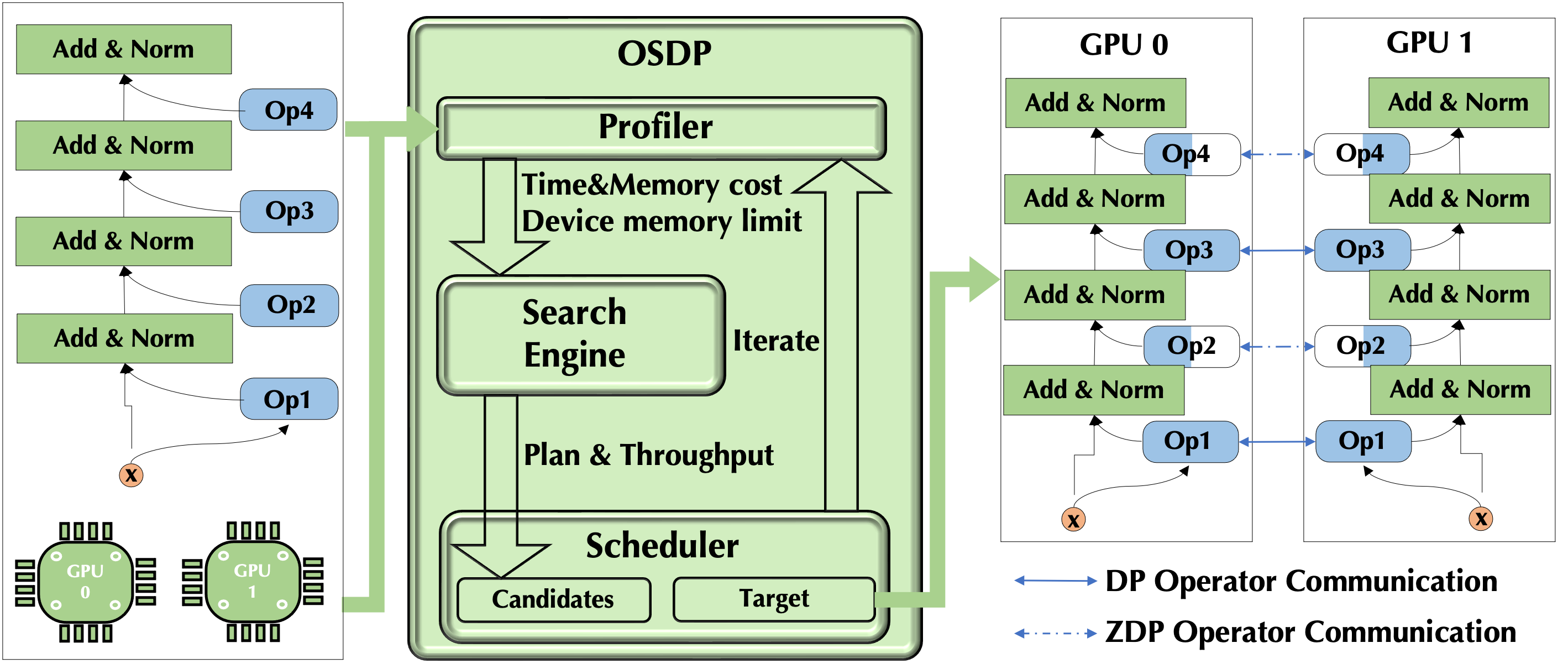}
\caption{Workflow of OSDP.}
\label{fig:OSDPworkflow}
\end{figure}

Figure~\ref{fig:OSDPworkflow} demonstrates the workflow of OSDP, which consists of three major modules: the \textit{Profiler}, the \textit{Search Engine}, and the \textit{Scheduler}. As shown in Algorithm~\ref{alg:CPSA}, these modules work together to maximize the overall system throughput adaptively and automatically. Below we introduce the workflow in depth.

\textbf{\textit{Profiler.}}
The \textit{Profiler} is responsible for estimating the memory and time cost. In each time of profiling, the \textit{Search Engine} suggests an execution plan $\boldsymbol{p}$ and a batch size $b$ (along with the model description and device information provided by users). The \textit{Profiler} follows the cost model as discussed in Section~\ref{sec:formulation} and outputs the estimated memory and time cost.
% Given the model description, device information, training batch size, and the candidate execution plan, it follows the cost model as discussed in Section~\ref{sec:formulation} and outputs the estimated memory and time cost.

\textbf{\textit{Search Engine.}}
The \textit{Search Engine} takes as input the memory cost and time cost estimated by the \textit{Profiler}, and adopts Depth First Search (DFS) as the search method. DFS traverses and makes decisions on each operator in the model based on their time and memory cost, ensures the generated plan minimizes the overall time cost while the memory cost is under device memory limit, and eventually, outputs the optimal execution plan and its corresponding estimated system throughput. Additionally, two intuitive pruning schemes are introduced --- if the current memory usage exceeds memory limit or the current time cost exceeds the best plan so far, we will prune the searching immediately. Eventually, it takes merely 9-307 seconds in our experiments to complete the search process, which is worthy as the improvement in efficiency can save hours or even days in large model training.

\textbf{\textit{Scheduler.}}
The \textit{Scheduler} iteratively collects the output plan and throughput from \textit{Search Engine} as candidates, and increases the training batch size output to \textit{Profiler} until the minimum possible overall memory cost exceeds device memory limit. Normally when memory is sufficient, a larger training batch size demonstrates a higher system throughput, we only need to output the last candidate of \textit{Scheduler} as the overall optimal execution plan. However, OSDP makes full use of the device memory in every batch size training, which makes it possible to train with a smaller batch size to get a higher system throughput. In this case, \textit{Scheduler} chooses the plan with the highest estimated system throughput among all the candidates and outputs it as the final execution plan.

\begin{figure*}[!t]
\begin{minipage}{.6\textwidth}
    \centering
    \includegraphics[width=\linewidth]{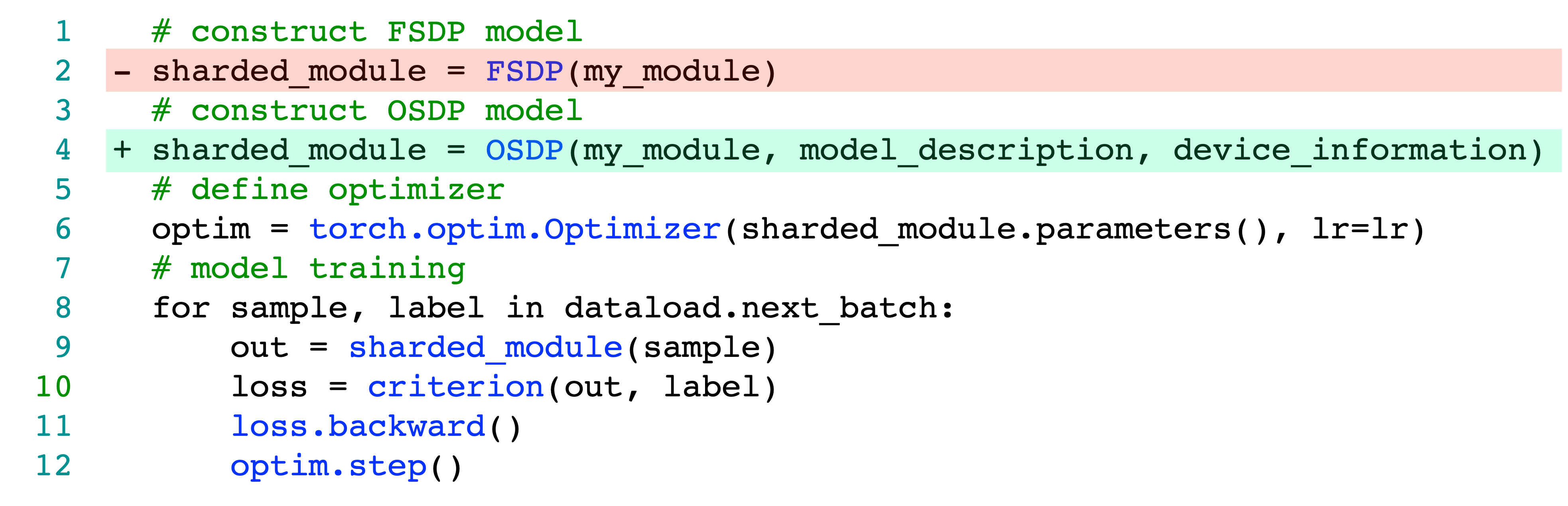}
    \captionof{figure}{Comparison between FSDP and OSDP.}
    \label{fig:OSDP_API}
\end{minipage}
% \begin{minipage}{.01\textwidth}
% $ $
% \end{minipage}
\begin{minipage}{.4\textwidth}
    \centering
    \includegraphics[width=\linewidth]{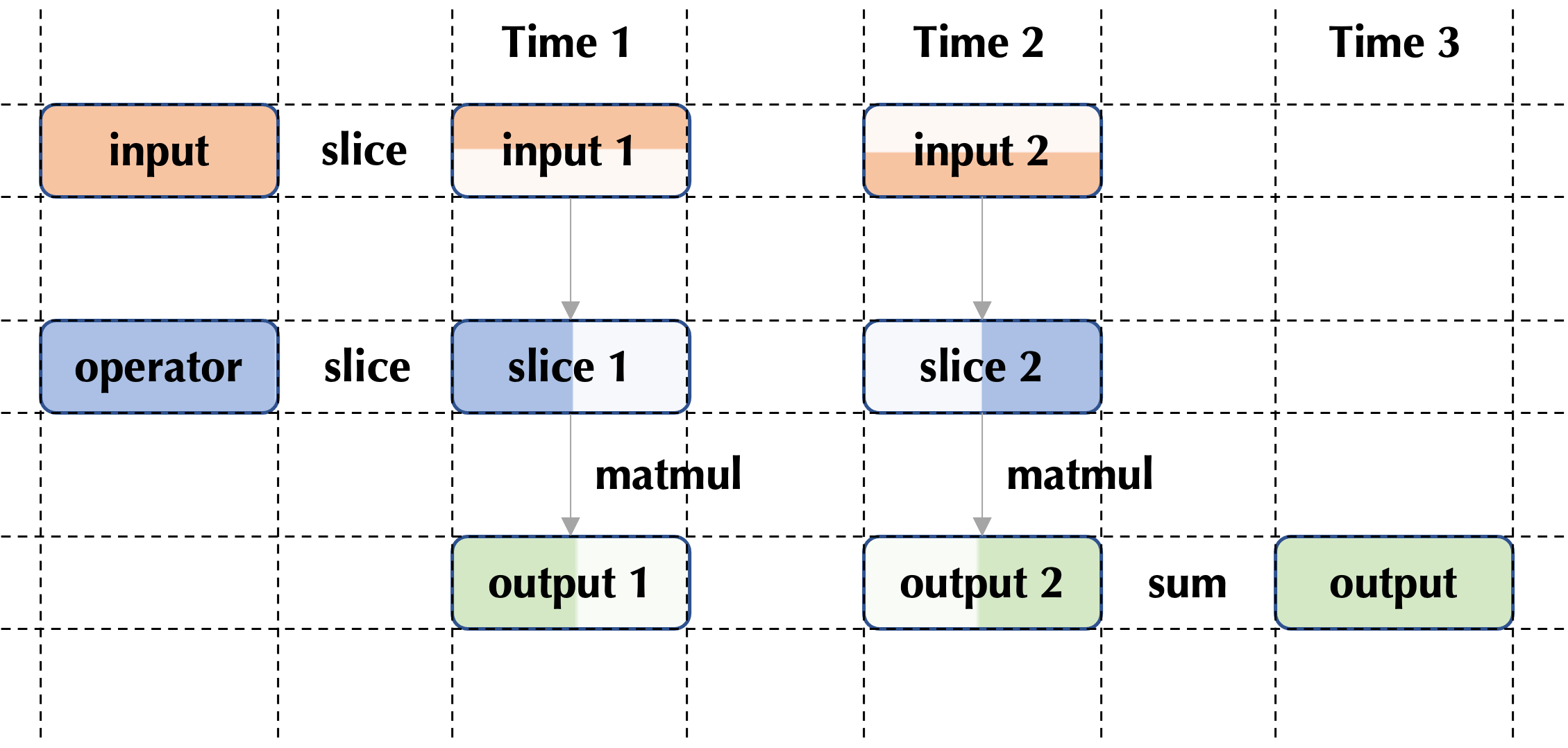}
    \captionof{figure}{Workflow of operator splitting.}
    \label{fig:operatorsplittingworkflow}
\end{minipage}
\end{figure*}

% \begin{figure}[!t]
% \centering
% \includegraphics[width=1.0\linewidth]{figure/OSDP_API.png}
% \caption{Comparison between FSDP and OSDP.}
% \label{fig:OSDP_API}
% \end{figure}

\subsection{Implementation and Optimization}
\label{sec:impl}

\textbf{\textit{Implementation.}}
We implement OSDP on top of PyTorch and FairScale. To be specific, OSDP is designed as an efficient distributed training framework that adaptively and automatically determines the data parallel mechanisms. Furthermore, to be user-friendly, we provide a simple but convenient API interface following the FSDP module in FairScale. As shown in Figure~\ref{fig:OSDP_API}, by only modifying a few lines of codes, users can easily switch from FSDP to OSDP for better efficiency and scalability.

% \begin{figure}[!t]
% \centering
% \includegraphics[width=1.0\linewidth]{figure/OSAworkflow.png}
% \caption{Workflow of operator splitting.}
% \label{fig:operatorsplittingworkflow}
% \end{figure}

\textbf{\textit{Operator Splitting.}}
In order to alleviate peak memory footprints brought by the huge, bottleneck operators, we further introduce the operator splitting scheme to cooperate with OSDP. Specifically, existing ZeRO-based systems (such as DeepSpeed and FairScale) shard model states across workers during model initialization stage and re-gather them during training stage.
Obviously, there is a memory surge during the gathering process ---
for the huge operators (e.g., \texttt{MatMul} operators with large hidden sizes), gathering the corresponding gigantic tensors could turn into peak memory usage and might exceed the device memory.
For instance, one of the  \texttt{MatMul} operators in the GPT-3 model contains 0.6 billion parameters, which ends up to consume 2.24 GB of memory. 

The operator splitting aims at reducing the impact of such gigantic tensors in OSDP.
Intuitively, given a huge \texttt{MatMul} operator, the essential idea is to split the model into several slices and sequentially process them. By doing so, the memory consumed by different slices can be released serially so that the peak memory can be reduced greatly. Figure~\ref{fig:operatorsplittingworkflow} illustrates the workflow, including three steps. First, both the last dimension of the input data and the first dimension of the operator are partitioned into multiple slices according to an artificially determined slice granularity. Then, the computation of each slice is executed sequentially. Finally, all computation results are summed as the final output.

Combined with ZDP, operator splitting amortizes the memory from $size(\text{\texttt{MatMul}})$ to $\frac{size(\text{\texttt{MatMul}})}{{slice}\_{granularity}}$, which is extremely beneficial for large-scale models. In addition, since our distributed training deployment supports the overlapping between computation and communication \cite{pytorchdistributed}, as long as the communication cost remains a system bottleneck, the computational overhead caused by slicing and summation can be completely hidden. Thus, the extra overhead of operator splitting on the overall training time is almost negligible.

Putting them together, the performance of OSDP can be improved with operator splitting. As operator splitting partitions one operator into multiple slices, OSDP can treat each slice individually. For instance, instead of assigning the DP or ZDP mode to an entire operator, OSDP can first partition the operator into 4 slices and process 1 of them in the ZDP mode and 3 of them in the DP mode. By such means, OSDP can provide a variety of choices for a single operator and search for a more fine-grained execution plan for the model. 
% Meanwhile, as mentioned in Section~\ref{sec:OP}, operator splitting minimizes memory surge in training, which provides OSDP with the ability to undertake a larger batch size and further optimize the system throughput.

% \input{approach}

% \input{plot/asdps_throughput.tex}
% \input{plot/checkpoint_throughput.tex}
\section{Experiments}
\label{sec:exp}

\subsection{Experimental Setup} \label{sec:tasksetting}
We first introduce the experimental setup in this work.

\textbf{\textit{Environments.}}
We conduct experiments on two types of hardware environments. Most of our experiments are performed on a laboratorial server equipped with 8 NVIDIA RTX TITAN 24 GB GPUs using PCIe 3.0. We mainly use this hardware environment to assess the effectiveness and efficiency of OSDP. For the multi-server experiments, two cloud servers equipped with NVIDIA A100 GPUs are utilized. The network bandwidth between the two servers is 100 Gb.

\textbf{\textit{Baselines.}}
We compare OSDP with both pure and hybrid parallel strategies. For pure parallel strategies, we choose PyTorch DDP, GPipe, and Megatron-LM \cite{pytorchdistributed,huang2019gpipe,shoeybi2020megatronlm} as the representatives of DP, PP, and TP, respectively. FairScale~\cite{baines2021fairscale} is chosen as the representative of FSDP/ZeRO. OSDP-base represents OSDP without operator splitting. We also conduct experiments on hybrid parallelism. To be specific, we compare with DeepSpeed 3D parallelism \cite{rasley2020deepspeed}, which integrates DP, PP, and TP together. In addition, since OSDP can be regarded as a substitute of DP, we further replace the DP dimension in 3D parallelism to form a new hybrid parallel strategy called 3D+OSDP, which demonstrates the compatibility of OSDP with existing hybrid strategies. To achieve a fair comparison, we tune the combinations of parallel strategies for hybrid parallelism and report the one with the best performance. By default, we set the slice granularity of our operator splitting technique as 4, and we will conduct more experiments with varying granularities in Section~\ref{sec:more_exp}.

\begin{table}[t]
\centering
\small
\caption{Statistics of Models}
\label{tab:model_config}
\vspace{-2mm}
\scalebox{0.65}{
\begin{tabular}{c|ccccc}
\toprule 
\makecell{Model} & Layer Num  & Operator Num & Hidden Size & Param. Num\\
\midrule 
N\&D & 48-96  & 98-194 & 1024-1536 & 1.3-2.9B\\
W\&S & 2-4  & 6-10 & 6144-12288 & 1.7-4B\\
I\&C & 24-96  & 50-194 & 1024-4096 & 0.9-2.3B\\
\bottomrule
\end{tabular}}
\vspace{-4mm}
\end{table}

\textbf{\textit{Models.}}
We choose minGPT\footnote{\url{https://github.com/karpathy/minGPT}} as our experimental model base, which is a well-known PyTorch re-implementation of GPT training. As shown in Table~\ref{tab:model_config}, We propose three different types of models: narrow \& deep (\textbf{N\&D}) models, wide \& shallow (\textbf{W\&S}) models, and inconsistent \& consecutive models (\textbf{I\&C}) models. N\&D models have numerous layers with small hidden sizes, which represent models such as GPT-2, Bert, and T5 \cite{radford2019language,devlin2018bert,raffel2019exploring}. W\&S models have few layers with large hidden sizes, which represent models such as GPT-3 \cite{brown2020language} that can only place part of its layer on one device. I\&C models have layers with different hidden sizes, which represent models such as Swin transformer \cite{swintransformer}. 
For each type of models, we conduct experiments with several configurations to evaluate the universal applicability of our work. All experiments are executed for 100 iterations and the averaged statistics are reported.

\begin{figure*}[!t]
\centering
\includegraphics[width=1\linewidth]{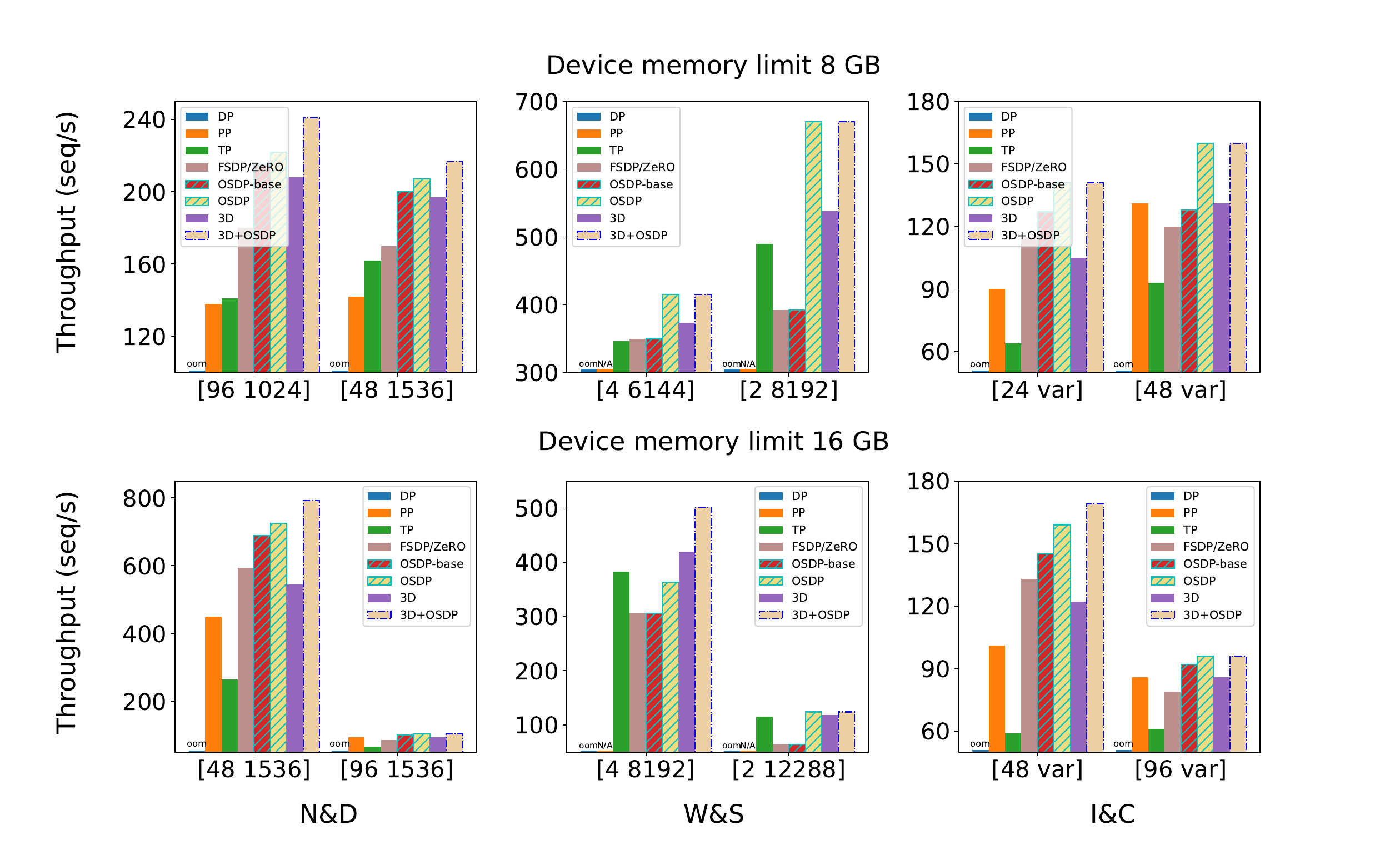}
\caption{End-to-end comparison of different parallel strategies with 8 GPUs. The x-axis represents different settings for the number of model layers and hidden sizes, the x label represents the model type, and the y-axis represents the overall training throughput. ``OOM'' indicates out of memory and ``N/A'' indicates not applicable (PP requires at least 8 layers, so it is not applicable on W\&S models).}
\label{fig:8gpu_ex}
\end{figure*}

\begin{figure}[!t]
\centering
\includegraphics[width=1\linewidth]{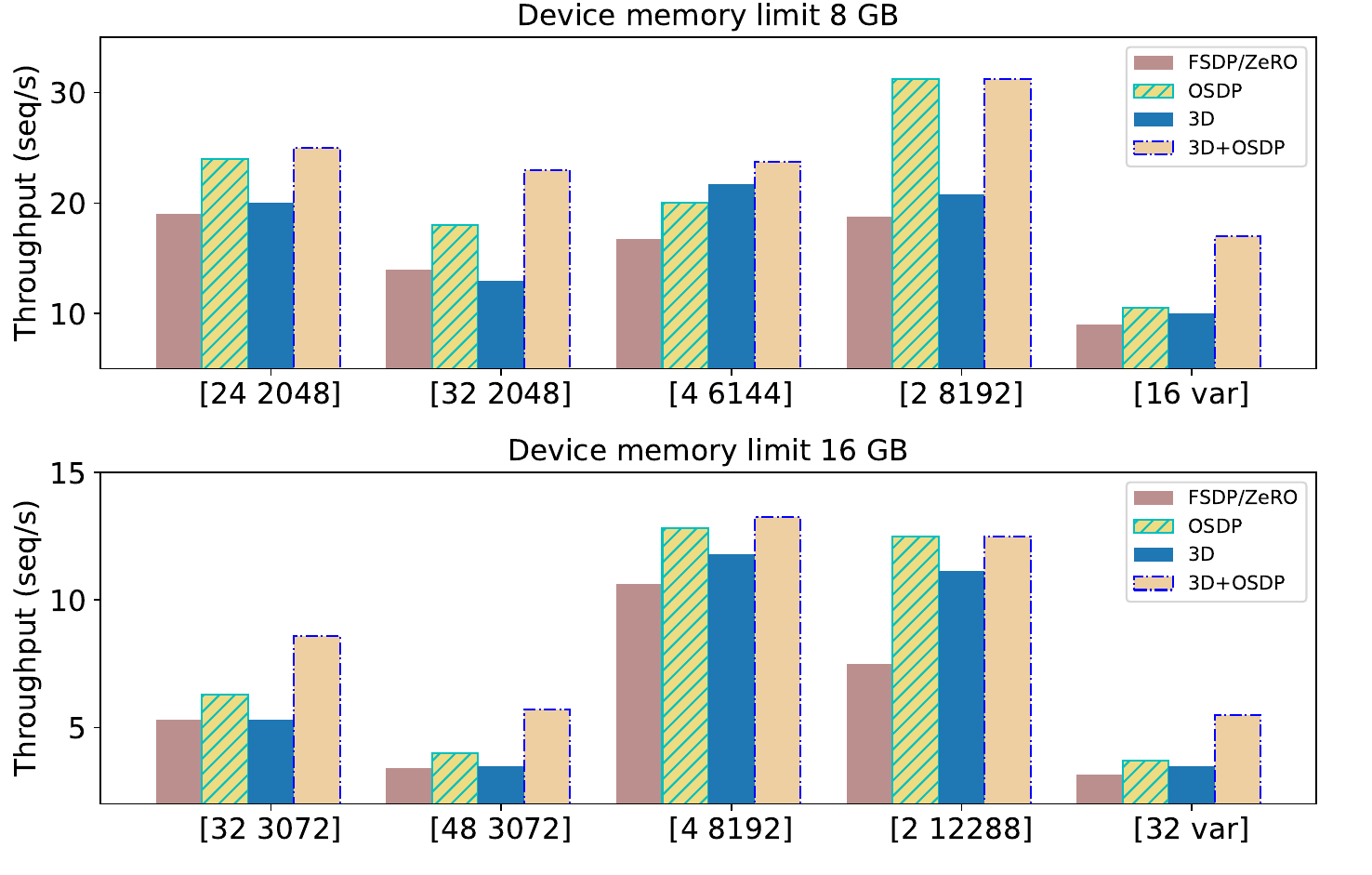}
\caption{End-to-end comparison of different parallel strategies with 16 GPUs.}
\vspace{-4mm}
\label{fig:16gpu_ex}
\end{figure}

\subsection{End-to-end Comparison}
\label{sec:e2e_exp}

We first compare the overall training throughput of all counterparts by conducting experiments on N\&D, W\&S and I\&C under the GPU memory limit of 8G and 16G, respectively. The results are provided in Figure~\ref{fig:8gpu_ex} and Figure~\ref{fig:16gpu_ex}.

% As shown in \autoref{fig:8gpu_ex}, experimental results on N\&D demonstrate that OSDP outperforms pure strategies on all tasks, a maximum of 2x acceleration is achieved , an average of 17.2\% speedup is achieved on all N\&D tasks, and operator splitting increases the acceleration effect of OSDP to a maximum of 23\% and an average of 22\% by providing a finer-grained implementation as mentioned in \autoref{sec:OSDP+OP}. In addition, experimental results on W\&S demonstrate that OSDP can hardly achieve promising acceleration effect relative to FSDP on models with few layers and large hidden sizes, while operator splitting alleviates peak memory footprint of OSDP and enables it with larger batch size training, and as a result a maximum of 92\% acceleration is achieved on model with 2 layers and hidden size 12288, an average of 50\% speedup is achieved on all W\&S tasks. Moreover, experimental results on I\&C shows that OSDP realizes a maximum and an average of 16\% and 10\% accelerations among all I\&C tasks, and operator splitting promotes them to 33\% and 24\% respectively.

\begin{figure*}
\begin{minipage}[t]{0.24\textwidth}
\centering
\begin{tikzpicture}
    \begin{axis}[
        legend style={at={(0.8,1.25)},anchor=north, nodes={scale=0.5, transform shape}},
        xlabel={Slice granularity},
        ylabel={Time cost (s)}, 
        symbolic x coords={0, 4, 6, 8, 16},
        %xtick distance=1.2,
        xticklabels={},
        extra x ticks={0, 4, 6, 8, 16},
        extra x tick labels={0, 4, 6, 8, 16},
        ytick distance=0.01,
        ymajorgrids,
        scale only axis,
        width=0.65\linewidth,
        height=67pt,
        ymin=0,
        ylabel shift=-3pt,
        ylabel near ticks,
        %xlabel shift=-5.9pt,
        %xlabel style={at={(0.5,-0.150)}},
        xlabel near ticks,
        xtick pos=bottom,
        y axis line style={opacity=0},
        grid style=dashed,
    ]
    \addplot+[sharp plot, black] coordinates {
        (0, 0.027921486) (4, 0.035680008) (6, 0.036335516) (8, 0.038943005) (16, 0.059446621)
    };
    \addplot+[sharp plot, red] coordinates {
        (0, 0.032728004) (4, 0.037297058) (6, 0.039299631) (8, 0.046210623) (16, 0.056648779)
    };
    \legend{768, 1024}
    \end{axis}
\end{tikzpicture}
\subcaption{Operators with hidden size 768 and 1024.}
\end{minipage}
\begin{minipage}[t]{0.24\textwidth}
\centering
\begin{tikzpicture}
    \begin{axis}[
        legend style={at={(0.8,1.25)},anchor=north, nodes={scale=0.5, transform shape}},
        xlabel={Slice granularity},
        ylabel={Mem cost (MB)}, 
        symbolic x coords={0, 4, 6, 8, 16},
        %xtick distance=1.2,
        xticklabels={},
        extra x ticks={0, 4, 6, 8, 16},
        extra x tick labels={0, 4, 6, 8, 16},
        ytick distance=20,
        ymajorgrids,
        scale only axis,
        width=0.65\linewidth,
        height=67pt,
        ymin=0,
        ylabel shift=-3pt,
        ylabel near ticks,
        %xlabel shift=-5.9pt,
        %xlabel style={at={(0.5,-0.150)}},
        xlabel near ticks,
        xtick pos=bottom,
        y axis line style={opacity=0},
        grid style=dashed,
    ]
    \addplot+[sharp plot, black] coordinates {
        (0, 72.89) (4, 72.79) (6, 73.57) (8, 72.83) (16, 73.72)
    };
    \addplot+[sharp plot, red] coordinates {
        (0, 105.01) (4, 85.9) (6, 84.43) (8, 82.03) (16, 80.11)
    };
    \legend{768, 1024}
    \end{axis}
\end{tikzpicture}
\subcaption{Operators with hidden size 768 and 1024.}
\end{minipage}
\begin{minipage}[t]{0.24\textwidth}
\centering
\begin{tikzpicture}
    \begin{axis}[
        legend style={at={(0.8,1.25)},anchor=north, nodes={scale=0.5, transform shape}},
        xlabel={Slice granularity},
        ylabel={Time cost (s)}, 
        symbolic x coords={0, 4, 6, 8, 16},
        %xtick distance=1.2,
        xticklabels={},
        extra x ticks={0, 4, 6, 8, 16},
        extra x tick labels={0, 4, 6, 8, 16},
        ytick distance=1,
        ymajorgrids,
        scale only axis,
        width=0.65\linewidth,
        height=67pt,
        ymin=0,
        ylabel shift=-3pt,
        ylabel near ticks,
        xlabel near ticks,
        %xlabel shift=-5.9pt,
        %xlabel style={at={(0.5,-0.150)}},
        xtick pos=bottom,
        y axis line style={opacity=0},
        grid style=dashed,
    ]
    \addplot+[sharp plot, black] coordinates {
        (0, 1.855350685) (4, 1.846748638) (6, 1.85100131) (8, 1.8698493) (16, 1.848715687)
    };
    \addplot+[sharp plot, red] coordinates {
        (0, 5.391268492) (4, 5.160170031) (6, 5.202160311) (8, 4.8147048) (16, 4.881180573)
    };
    \legend{8192, 12288}
    \end{axis}
\end{tikzpicture}
\subcaption{Operators with hidden size 8192 and 12288.}
\end{minipage}
\begin{minipage}[t]{0.24\textwidth}
\centering
\begin{tikzpicture}
    \begin{axis}[
        legend style={at={(0.8,1.25)},anchor=north, nodes={scale=0.5, transform shape}},
        xlabel={Slice granularity},
        ylabel={Mem cost (MB)}, 
        symbolic x coords={0, 4, 6, 8, 16},
        %xtick distance=1.2,
        xticklabels={},
        extra x ticks={0, 4, 6, 8, 16},
        extra x tick labels={0, 4, 6, 8, 16},
        ytick distance=2000,
        ymajorgrids,
        scale only axis,
        width=0.65\linewidth,
        height=67pt,
        ymin=0,
        ylabel shift=-3pt,
        ylabel near ticks,
        %xlabel shift=-5.9pt,
        %xlabel style={at={(0.5,-0.150)}},
        xlabel near ticks,
        xtick pos=bottom,
        y axis line style={opacity=0},
        grid style=dashed,
    ]
    \addplot+[sharp plot, black] coordinates {
        (0, 4616.4) (4, 3080.85) (6, 2571.03) (8, 2313.39) (16, 2311.09)
    };
    \addplot+[sharp plot, red] coordinates {
        (0, 10380.59) (4,6925.39) (6, 5773.95) (8, 5197.97) (16, 5193.06)
    };
    \legend{8192, 12288}
    \end{axis}
\end{tikzpicture}
\subcaption{Operators with hidden size 8192 and 12288.}
\end{minipage}
\caption{Figure(a)-(b) demonstrate the impact of operator splitting on operators with small hidden sizes, and (c)-(d) present the impact of operator splitting on operators with large hidden sizes (8 GPUs).}
\label{fig:sg_1234}
\end{figure*}
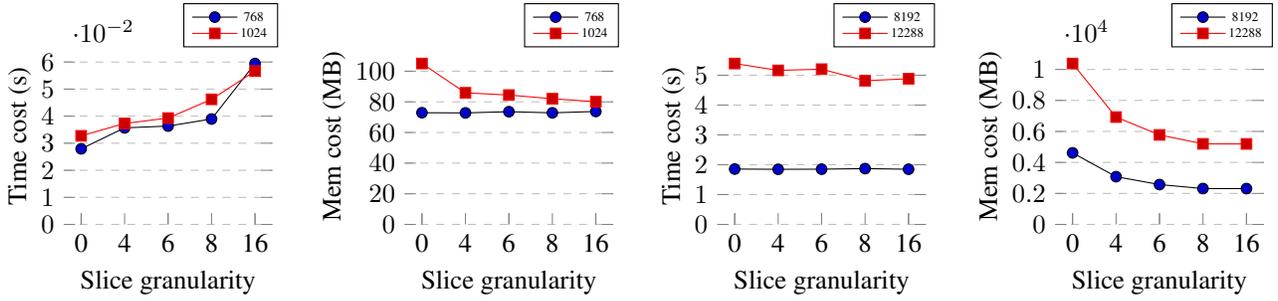

\begin{figure*}[!t]
\begin{minipage}{.48\textwidth}
    \centering
    \includegraphics[width=\linewidth]{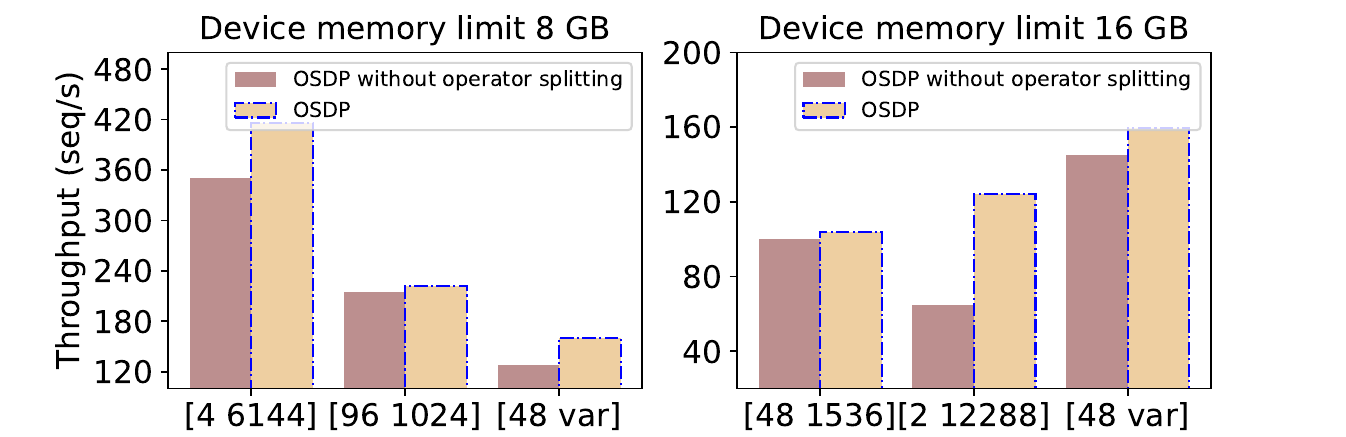}
    \captionof{figure}{Throughput comparison of OSDP with and without the operator splitting technique (8 GPUs).}
    \label{fig:os}
\end{minipage}
\begin{minipage}{.01\textwidth}
$ $
\end{minipage}
\begin{minipage}{.48\textwidth}
    \centering
    \includegraphics[width=\linewidth]{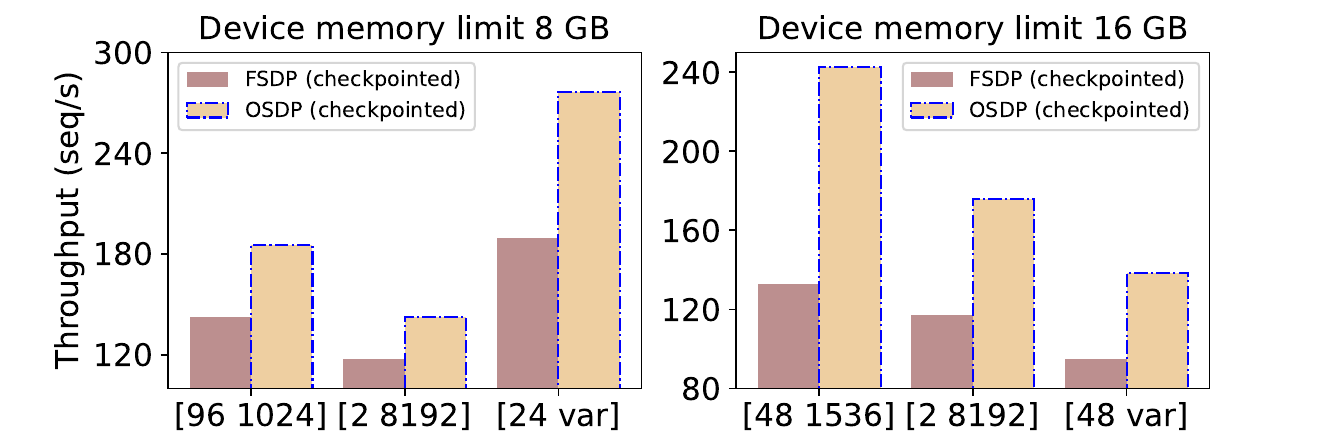}
    \captionof{figure}{Throughput comparison of OSDP and FSDP with the checkpointing technique (8 GPUs).}
    \label{fig:checkpoint}
\end{minipage}
\end{figure*}

\textbf{\textit{Comparison with Pure Parallelism.}}
We first discuss the empirical results of pure parallel strategies (i.e., DP, PP, TP, FSDP, and OSDP). 

As shown in Figure~\ref{fig:8gpu_ex}, on the N\&D tasks, OSDP achieves a maximum of 174\% acceleration compared with the other pure parallel strategies. In particular, OSDP outperforms FSDP with a maximum and an average of 23\% and 22\% speedup, respectively. It verifies that fine-grained memory management method of OSDP is able to make a balance between the memory consumption and training efficiency, and therefore, provide the system with a higher end-to-end training throughput. On the W\&S tasks, due to the huge size of operators, ZeRO optimizer is unsuitable for such a type of models, which leads to the unsatisfactory performance of FSDP. However, since OSDP alleviates the peak memory footprint by splitting the huge operators and making fine-grained execution plans, it exibits a much better performance than FSDP --- compared with the pure parallel counterparts, OSDP achieves a maximum of 92\% and an average of 32\% speedup, respectively. Finally, on the I\&C tasks, OSDP achievies a maximum of 168\% and an average of 33\% speedup, demonstrating its flexibility.

The results of two-server experiments also verify the ability of OSDP. As shown in Figure~\ref{fig:16gpu_ex}, OSDP outperforms FSDP by a maximum of 67\% and an average of 29\%.

\textbf{\textit{Comparison with Hybrid Parallelism.}}
In addition to pure parallelism, recent studies have proved that using combinatorial parallel strategies could bring further improvement to large-scale model training. In order to assess the performance of OSDP interacting with other parallelism, we further incorporate OSDP with TP and PP to obtain a stronger hybrid parallel strategy, i.e., 3D+OSDP. As shown in Figure~\ref{fig:8gpu_ex} and Figure~\ref{fig:16gpu_ex}, 3D+OSDP consistently achieves the highest training throughput in all experiments. In short, 3D+OSDP outperforms DeepSpeed 3D parallelism by a maximum of 73\% and an average of 31\%, and achieves a maximum of 184\% and an average of 38\% acceleration compared with the other baselines. These empirical results prove that OSDP can well fit with other parallel strategies, and therefore, make the training system more flexible and universally applicable.

\subsection{More Experiments}
\label{sec:more_exp}

\textbf{\textit{Effectiveness of Operator Splitting.}}
To evaluate the effectiveness of the operator splitting technique, we conduct experiments to investigate its impact on memory and time cost, and assess the improvement it contributes to the overall training efficiency.

We first evaluate the impact of operator splitting on ZDP training. To be specific, we conduct experiments on operators with hidden sizes from 1024 to 12288 and vary the slice granularity from 0 to 16 (a slice granularity of 0 indicates no operator splitting is performed). As shown in Figure~\ref{fig:sg_1234}, operator splitting alleviates peak memory footprints during training in all experiments, and a maximum of 50\% reduction in memory cost is observed. In addition, for operators with small hidden sizes (i.e., 768 and 1024), larger slice granularity leads to an increase in the time cost, while for operators with large hidden sizes (i.e., 8196 and 12288), smaller slice granularity cannot fully reduce the memory cost. These empirical results demonstrate that by tuning the slice granularity for different models and even different operators, OSDP is able to achieve better training performance. And it is an interesting topic to explore how to automatically suggest a desirable slice granularity to each operator according to the model description and device information.

% \blue{Therefore, in OSDP tasks with operator splitting, we dynamically select the slice granularity that fully optimizes operator's memory footprint while incurring acceptable additional time cost according to different hidden sizes.}

% \textbf{Operator splitting impact on memory and time cost} To evaluate the impact of operator splitting on zero redundancy data parallel training, we conduct multiple experiments on operators with hidden sizes from 1024 to 12288 and slice granularity from 0 to 16. As shown in \autoref{fig:sg_1234}, operator splitting alleviates peak memory footprints during training in all cases, a maximum of 50\% reduction in memory cost is witnessed. In addition, for operators with small hidden sizes (768, 1024), larger slice granularity leads to a surge in time cost, and for operators with large hidden sizes (8196, 12288), smaller slice granularity can not fully optimize the memory cost. Therefore, in OSDP tasks with operator splitting, we dynamically select the slice granularity that fully optimizes operator's memory footprint while incurring acceptable additional time cost according to different hidden sizes.

Next, we evaluate the overall training efficiency of OSDP with and without the operator splitting technique under the GPU memory limit of 8G and 16G, respectively. The results are shown in Figure~\ref{fig:os}. In N\&D, approximately 25\% of operators are partitioned using splitting for finer mode selection and higher throughput. In W\&S, all operators are partitioned, reducing peak memory and enabling larger batches, enhancing throughput. In I\&C, around 50\% of operators are partitioned, prioritizing larger ones and selectively partitioning smaller ones based on demand, maximizing throughput. In short, the operator splitting technique consistently improves the training throughput by 3\%-92\%.

\textbf{\textit{Integrating with Checkpointing.}}
As introduced in Section~\ref{sec:bg}, checkpointing is a widely used technique to eliminate the impact on memory brought by activations. Furthermore, checkpointing is usually integrated with the ZDP mode in many real-world applications \cite{rajbhandari2020zero,ren2021zerooffload,rajbhandari2021zeroinfinity}. To evaluate the impact of checkpointing, we compare OSDP and FSDP with checkpointing enabled. As shown in Figure~\ref{fig:checkpoint}, checkpointing increases the training throughput of OSDP and FSDP (compared with the results in Figure~\ref{fig:8gpu_ex}). However, the improvement on OSDP is larger --- when intergrating with checkpointing, OSDP achieves up to 108.3\% and an average of 52.9\% speedup compared with FSDP. In fact, when checkpointing is enabled with the ZeRO optimizer, the recomputation process before backward propagation requires an additional round of gathering since each GPU does not maintain a full copy of model states, which has a big impact on the overall training efficiency. By making fine-grained decisions, OSDP can leave operators with smaller memory overhead in DP mode, and therefore, mitigate the side-effect of checkpointing.

\section{Conclusion}
In this work, we proposed a novel automatic parallel system OSDP, which optimizes data parallel training by making fine-grained trade-offs between memory consumption reduction and the training efficiency improvement. In addition, OSDP supports operator splitting for more fine-grained execution plan decisions and memory optimization. Empirical results demonstrate that OSDP outperforms the state-of-the-art parallel training systems in multiple regards, and achieves up to $2.84\times$ of speedup in terms of the overall system throughput.

\appendix

\section*{Acknowledgments}
This work is supported by National Key R\&D Program of China (2022ZD0116315), National Natural Science Foundation of China (61832001,U22B2037), and PKU-Tencent joint research Lab. Fangcheng Fu and Bin Cui are the corresponding authors.

% \nocite{tinyscript,celuvfl,Miao_2022}
\nocite{tinyscript,celuvfl}
\bibliographystyle{named}

\bibliography{docx}

\end{document}